\documentclass[a4paper]{cas-sc} 
\usepackage{relsize}

\usepackage[authoryear]{natbib}
\usepackage{amsmath}

\usepackage{graphicx} 
\usepackage{subcaption}
\graphicspath{ {./figs/} }
\usepackage{float}
\usepackage{multirow}
\usepackage{rotating}
\usepackage{tabularx}
\usepackage{cleveref}

\def\tsc#1{\csdef{#1}{\textsc{\lowercase{#1}}\xspace}}
\tsc{WGM}
\tsc{QE}
\tsc{EP}
\tsc{PMS}
\tsc{BEC}
\tsc{DE}

\usepackage[acronym]{glossaries}
\newacronym{acis}{ACIS}{Accessibility Corridor Impact Score}

\begin{document}
\let\WriteBookmarks\relax
\def\floatpagepagefraction{1}
\def\textpagefraction{.001}

\shorttitle{Stress-testing Road Networks}

\shortauthors{Hannah Schuster et~al.}

\title [mode = title]{Stress-testing Road Networks and Access to Medical Care}                      

\tnotetext[1]{The authors acknowledge support from an Austrian Applied Research Agency (FFG) grant: CRISP (FFG No 887554).}

%
\author[1,2]{Hannah Schuster}[type=editor,
                        auid=000,bioid=1,
                        orcid= xx]

\cormark[1]

\fnmark[1]

\ead{schuster@csh.ac.at}

\author[1,2]{Axel Polleres}

\author[3,4,1]{Johannes Wachs}
\cormark[1]
\ead{johannes.wachs@uni-corvinus.hu}

\affiliation[1]{organization={Complexity Science Hub Vienna},
    addressline={Josefstädterstraße}, 
    city={Vienna},
    postcode={AT-1080}, 
    country={Austria}}
\affiliation[2]{organization={Vienna University of Economics and Business},
    addressline={Welthandelsplatz}, 
    city={Vienna},
    postcode={AT-1020},
    country={Austria}}

\affiliation[3]{organization={Corvinus University of Budapest},
    addressline={F\H{o}v\'am T\'er 8}, 
    city={Budapest},
    postcode={1093},
    country={Hungary}}

\affiliation[4]{organization={Centre for Economic and Regional Studies, E\"otv\"os Lor\'and Research Network},
    addressline={T\'oth K\'alm\'an u. 4.}, 
    city={Budapest},
    postcode={1097},
    country={Hungary}}


\begin{abstract}
This research studies how populations depend on road networks for access to health care during crises or natural disasters. So far, most researchers rather studied the accessibility of the whole network or the cost of network disruptions in general, rather than as a function of the accessibility of specific priority destinations like hospitals. Even short delays in accessing healthcare can have significant adverse consequences. We carry out a comprehensive stress test of the entire Austrian road network from this perspective. We simplify the whole network into one consisting of what we call accessibility corridors, deleting single corridors to evaluate the change in accessibility of populations to healthcare. The data created by our stress test was used to generate an importance ranking of the corridors. The findings suggest that certain road segments and corridors are orders of magnitude more important in terms of access to hospitals than the typical one. Our method also highlights vulnerable municipalities and hospitals who may experience demand surges as populations are cut off from their usual nearest hospitals. Even though the skewed importance of some corridors highlights vulnerabilities, they provide policymakers with a clear agenda. 

\end{abstract}




\maketitle

\section{Introduction}
\label{sec:intro}
Our dependence on road networks to access emergency medical care increases in two important ways during crises. Crisis events, such as natural disasters, can create increased demands for access to medical services by causing injuries directly. Additionally, these can also disrupt the functionality of these networks themselves, increasing the time it takes to get to a hospital. In acute cases, we know that delays can cause markedly worse medical outcomes for patients \citep{core:murata_association_2013, jena_delays_2017}. As climate change increases the frequency of severe weather events \citep{core:mukherji_synthesis_nodate}, which can extensively disrupt road networks, we need to better understand not only the abstract resilience of infrastructure such as road networks \citep{antoniou_statistical_2008} but also the specific weaknesses of these systems in terms of access to medical care.

Indeed, we can expect 
such weaknesses in road networks: especially in geographically rugged terrain, transportation infrastructure is expensive and highly constrained by physical barriers~\citep{rodrigue_transportation_2020}. Road networks are rightfully built with cost efficiency as a priority alongside robustness to periodic maintenance and disturbances. At the same time, these growing networks, like other complex systems, are known to be highly vulnerable to unanticipated shocks~\citep{carlson2002complexity}. Much like the banking sector, in which an unexpected financial insolvency can cause cascades of bankruptcies~\citep{battiston2016complexity,diem2020minimal}, local problems in road networks impact transportation through the whole system~\citep{hackl_risk_2019, goldbeck_resilience_2019}. Hospitals also face critical ``tipping-points'' - above a certain capacity they deliver significantly worse care \cite{kuntz2015stress}. Likewise, macro-scale medical care systems can also break down in the face of unexpected shocks \citep{lo2019quantification, kaleta2022stress}. Policymakers in all three domains: finance, transport infrastructure, and medical care are increasingly turning to stress tests to analyze their systems and pinpoint weaknesses.

Yet to date, little work has been done on how stresses and problems in one system can impact provision of services in another. Whatever the cause of a disruption, the complexity of these networks makes it difficult to predict the effect of one disruption on the functioning of the whole system. Given the significant potential coupling of risks in road transportation networks and access to and provision of medical care, we propose to develop a suitable stress test to examine how road network disruptions impact access to medical care. The aim of such a stress test is to highlight critical road segments or corridors that provide access to medical care, population centers at risk of being cut off from care, and hospitals that may see sudden surges in demand during crises. 

We implement this stress test by applying simulation analysis to data on road and healthcare infrastructure in Austria. Simulation analysis has proven an effective tool in modeling relative risks and the importance of components of complex systems ~\citep{core:liu_network_2022,hackl_estimating_2018, core:van_ginkel_will_2022}. Elements of these systems highlighted by stress tests are natural candidates for resources and attention from planners.

Our stress test of road networks and their provision of access to medical care presents three novel aspects. First, we develop a measure to quantify access from population centers to medical care. Most quantitative work on the resilience of transportation systems to date focuses on the impact of disruptions by determining the cost of disruption or by measuring the change of accessibility of the whole system during specific scenarios. However, during a disaster, changes in global accessibility or costs may be of minor concern compared to the specifics of which roads are used in the provision of essential services like healthcare or fire protection. We know that small differences in travel time to emergency care can have a significant impact on mortality and other patient outcomes~\citep{core:murata_association_2013}. To this end, we modify an existing measure of accessibility in road networks~\citep{core:martin_assessing_2021} in order to classify the importance of links in a network relating to the accessibility of municipalities to the closest hospitals. 

A second challenge in stress testing the resilience of road networks at the scale of a whole country is their size, which can make an exhaustive calculation and comparison of outages and their consequences intractable. We, therefore, introduce and stress test a coarse-grained simplification of the road system: we merge road segments connecting municipalities to create a backbone representation of the Austrian road network. We can stress test this network more extensively and show that derived insights can be transferred to the more realistic fine-grained system.

A third contribution of our approach is that we quantify the impact of our stress tests along three orthogonal dimensions. We measure how road network disruptions limit people's access to hospitals, suggesting vulnerability of population centers. We quantify road importance by observing the effects of their disruption. Finally, we measure the vulnerability of hospitals to sudden surges in the population they are the first point of care for. Thus our framework provides multi-level insight. We note that our approach can be applied and generalised to both other countries (depending on data availability) or to the provision of other services in crisis situations (for instance, firefighting facilities).

In the remainder of this paper, we first review the related literature on road network resilience and access to emergency medical care (\Cref{sec:litreview}). We then introduce the case of the Austrian road network and relevant datasets,  and describe the methods and measures used to study (\Cref{sec:Overview}). We present and interpret the results of our stress tests in \Cref{sec:results}. Finally, we conclude by discussing our method, including its limitations and avenues for future work in \Cref{sec:concl}.

\section{Literature Review}
\label{sec:litreview}
During crisis events impacting entire regions, the accessibility of medical care is crucial, given its potential to influence patient outcomes. Indeed, there is ample evidence of a direct effect of the travel distance to a hospital and the mortality of patients. A study using a national database in Japan concluded that the ambulance distance to hospitals significantly correlates with macro-regional mortality risks for particula acute diseases such as acute myocardial infarction and brain infarction~\citep{core:murata_association_2013}. Consequently, also Planned road closures and infrastructure disruptions result in worse mortality outcomes: for instance, previous work found a sharp increase in acute myocardial infarction or cardiac arrest hospitalizations among Medicare beneficiaries in $11$ U.S. cities during major marathons \citep{jena_delays_2017}.

In summary, the accessibility of emergency medical care depends crucially on road networks, which are also especially vulnerable to environmental perturbations such as extreme weather events~\citep{BIL201590}. Therefore, as numerous studies have shown that events like heatwaves, heavy rainfall, droughts, and tropical weather cyclones have become more frequent and intense globally since the 1950s ~\citep{core:mukherji_synthesis_nodate}, the vulnerability of road networks is likely to increase. In addition, the problem of transportation networks and accessibility is especially salient in geographic areas with rugged terrain~\citep{rodrigue_transportation_2020} (as we face it for instance in alpine regions in Austria), since such conditions limit possible cost-efficient redundancies that would make such networks more robust. 

More generally, growing systems like road networks are known to be vulnerable to unanticipated shocks \citep{carlson2002complexity}. Indeed, there is a whole literature analyzing the resilience of networks that tend to function well in ``normal'' times but can fail catastrophically during unexpected disruptions. Researchers have begun to stress test these systems to analyze their weak points, where simulation analysis has demonstrated its efficacy in modeling relative risks and the importance of components of complex systems~\citep{core:liu_network_2022, MATTSSON201516}. Particular applications of these methods include financial markets~\citep{battiston2016complexity}, food suppliers~\citep{schueller_propagation_2022}, regional economies \citep{toth2022technology}, ride-sharing systems \citep{bokanyi2020understanding}, and software systems \citep{schueller2022modeling}. Here, the resilience of a network is generally determined by monitoring the response of systems to the cumulative elimination of sections according to random order, deterministic order of criticality, and deterministic order in areas at high risk~\citep{core:martin_assessing_2021}. Insights gained through stress tests can be used to help guide planning resources and areas of attention, in order to improve the robustness of diverse networks while their functionality remains unchanged~\citep{schneider_mitigation_2011, lin_location_2023}. 

Methodological approaches to measuring resilience of road networks, in particular, vary from quantifying the travel cost of a disruption~\citep{jenelius_importance_2006, xie_disrupted_2023} to quantifying the risk to the overall network~\citep{hackl_estimating_2018}. The healthcare system has specifically been studied from this perspective, especially since the Covid-19 Pandemic: the additional stress of the pandemic shed light on the various problems of healthcare systems. Stress tests of hospital networks and networks of doctors have demonstrated that macro-scale medical care systems can also breakdown in the face of unexpected shocks~\citep{lo2019quantification, kaleta2022stress}, manifesting, for instance, in a sudden surge in patients which may overwhelm individual hospitals~\cite{kuntz2015stress}. 

Despite the apparent interest, little work has been done to understand how the infrastructure that provides access to care is vulnerable to shocks. Therefore in the present paper, we concentrate on how road closures change patient flows and volumes to hospitals. While previous work has studied how transport infrastructure ensures the provision of essential goods to communities~\citep{core:Grocery_2020,anderson_underestimated_2022}, to the best of our knowledge, access to healthcare has not been considered from this perspective thus far.

\section{Data and Methods}
\label{sec:Overview}
We now outline the data and methods we will use to stress test the Austrian road system. Our aim is to evaluate the importance of specific parts of the network in terms of the population's access to healthcare at hospitals. We first describe how we create an abstracted network representation of the road system. We call the edges or links in this resulting network \textit{corridors} and define measures of corridor importance. We then outline the methodology of two kinds of stress tests we will apply to this network. Finally, we describe how to measure the impact of these tests on hospitals.
\subsection{Constructing a network of corridors}
There are many possible ways to represent a nationwide transport network. Our goal is to create a representation of the Austrian road network that is simplified enough so that extensive stress tests are computationally feasible and still fine-grained enough to capture important details. We begin with data from GIP (Graphenintegrations-Plattform) \footnote{https://www.gip.gv.at/} - an extensive open data source of Austrian transportation infrastructure segments, from hiking trails to highways and railroads. As we are interested in emergency response, we focus on roads that can be accessed via automobile.

\begin{figure}[h]
	\centering
	\includegraphics[width=1\textwidth]{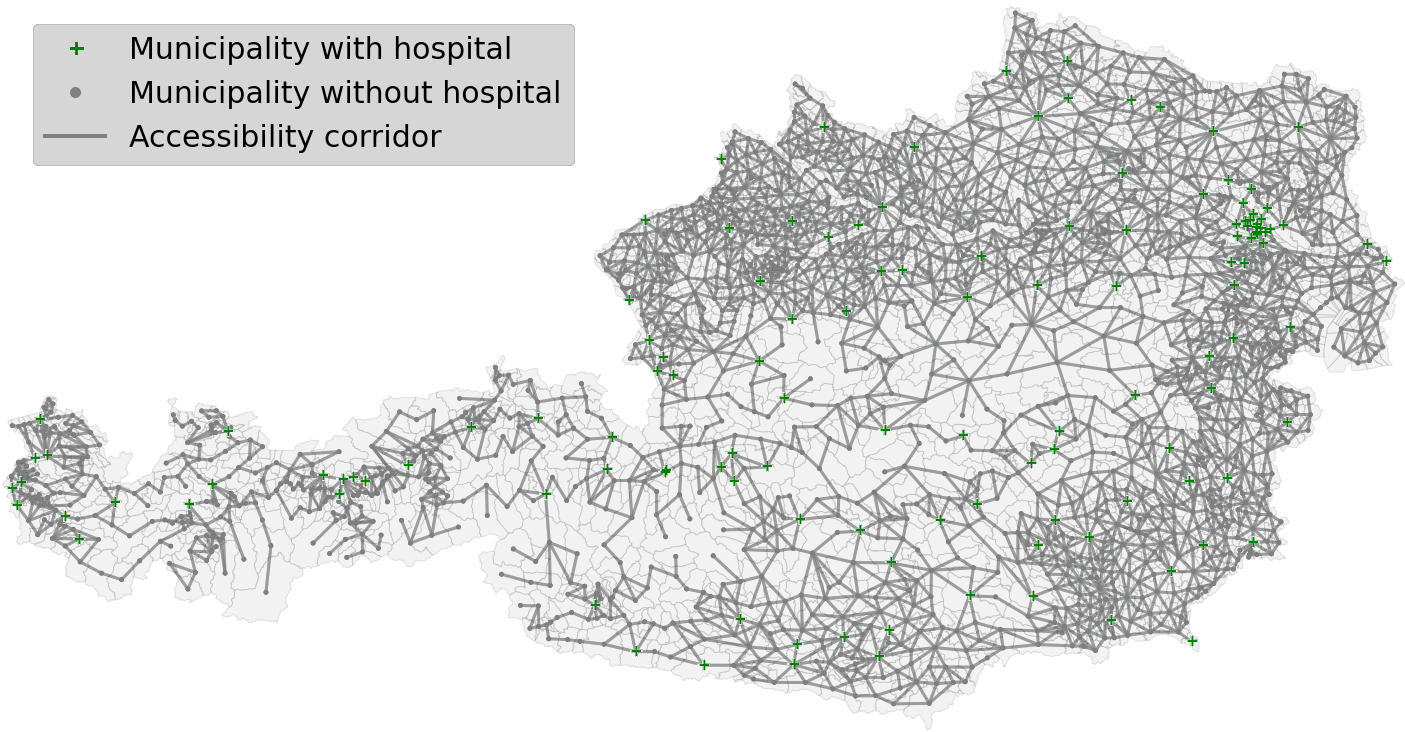}
	\caption{A coarse-grained representation of the Austrian road transportation system as a network. Nodes are municipalities and edges are accessibility corridors connecting them. Municipalities are colored and marked with a green cross if they contain a hospital.}
	\label{fig:coarse_grained_map}
\end{figure}

We first create a network of all roads, in which nodes are intersections of road segments and edges are roads. This is a very fine-grained representation of the system: with close to $1.5$ million links and about $1.3$ million nodes. Obviously, such a fine-grained representation presents a computational problem for network analysis and simulation: as we aim to simulate the removal of road segments and measure the impact on shortest paths to hospitals many times over, we therefore derive a coarse-grained abstraction of this network to keep shortest path calculations tractable while maintaining its core features. In the derived network nodes are \emph{municipalities} connected by an edge if there is a road segment ending in both municipalities. In other words: two municipalities are connected in the network if there is a direct path between them. We call these edges \textbf{accessibility corridors} or corridors for short, as they represent an abstraction of road connections between municipalities. We also record how many real-world roads between the municipalities are combined in a single corridor. This information is later used in the analysis section to emphasize the importance of the corridor in a real-world context. The resulting network representation of Austria is visualized in Figure~\ref{fig:coarse_grained_map}, with municipalities hosting a hospital highlighted.

\subsection{Measures of corridor importance}
\label{subsec:assessment_method}
Given our network representation of the Austrian road transportation network, we would like to quantify the importance of specific access corridors. The literature presents several ways to measure the importance of corridors and the impact of their closure on the movement of people, in general ~\citep{jenelius_importance_2006, xie_disrupted_2023, hackl_estimating_2018, core:Grocery_2020,anderson_underestimated_2022}. Our research introduces a new method by taking the accessibility of critical infrastructure into account when measuring importance. Specifically, we observe the impact of corridor closures on the accessibility of a municipality to its \emph{closest} hospital. Whether a corridor's closure causes a population to take a longer, indirect path to a hospital or forces them to go to a different hospital, we infer corridor importance from increases in travel times upon their removal weighted by the impacted population numbers. The changes in the accessibility measurement are used to implement a ranking of corridors, henceforth referred to as the \acrfull{acis}.

The \acrshort{acis} can be used to assess the impact of the initial stress test on the accessibility corridor network by estimating how a deletion impacts the distance to the closest hospital weighted by the impacted people. To determine the \acrshort{acis}, we start by calculating the integral of the cumulative population with respect to the distance for the baseline case and the stress-tested situation, using the trapezoid rule. Subsequently, the \acrshort{acis} can be determined by computing the difference between the baseline integral and the stressed integral, which can be expressed by the following formula:
\begin{equation}
	ACIS = \int_{0}^{dist_{max}} P_{base}(x) \,dx - \int_{0}^{dist_{max}} P_{stress}(x) \,dx,
\end{equation}
where $dist_{max}$ stands for the maximum distance between a hospital and a municipality in the original situation and the population functions $P_{base}(x)$ and $P_{stress}(x)$ characterize how many people have a hospital reachable within $x$ km.

\paragraph{Alternative Measures}
As an alternative to the \acrlong{acis} we also calculated a measure based on \cite{core:martin_assessing_2021}, where the authors introduce a measurement of a municipality's access to a full network of destinations based on the population distribution and the minimal distance to each node in the network. In our case, we modify this measure by switching the target variable to its closest hospital instead of all the other municipalities in the network given our focus on access to healthcare. The following equation represents our version of the accessibility measure, which we call the Hospital Accessibility of a municipality ($HA_{m}$):
\begin{equation}
	HA_m = max_{h \in H} \left(\frac{P_m}{d(m,h)}\right),
\end{equation}
where we measure the accessibility $HA_m$ of a municipality $m$ to the closest hospital by finding the maximum of the ratio of the population of the municipality $P_m$ divided by its distance $d$ to hospitals $h$ in Austria. The municipality's population is included to give greater weight to those municipalities with more people because the probability of someone needing a hospital increases with increasing population. 

To calculate an overall, aggregated accessibility measure for the entire country, which we call its Hospital Accessibility $HA$, we use the following formula:
\begin{equation}
	HA = \sum_{m \in M\backslash H}\frac{(HA_m * P_m)}{P_{M\backslash H}},
\end{equation}
where we calculate the sum of $HA_m$ over all municipalities $m$ in Austria, except for municipalities with a hospital, and then normalize each summand by a population factor $\frac{P_m}{P_{M\backslash H}}$, which takes the population $P_{M\backslash H}$ of all Austrian municipalities without a hospital into account. This measure captures the overall efficiency of the corridor network in terms of how well it gets people from population centers to hospitals.  

To rank the importance of the different corridors, we calculated the difference between the baseline accessibility score and the overall accessibility after stress testing the network. Specifically, if we remove corridor $k$, we define its impact $HA(k)$ as follows:
\begin{equation}
	HA(k) = \frac{HA_{baseline} - HA_{\backslash k}}{HA_{baseline}} *100,
\end{equation}
where $HA_{baseline}$ is the accessibility in the original situation and $HA_{\backslash k}$ is the overall hospital accessibility after accessibility corridor $k$ was deleted from the network.

As a third alternative besides the \acrshort{acis} and $HA$ measures of corridor importance, we also considered a popular way of ranking edges in networks called \textit{edge betweenness centrality}. In our context, this defines the importance of a corridor as follows:
\begin{equation}
	c_B (\text{corridor } e)= \sum_{s \in M,t \in H;  d(s,t)\leq 100km}\frac{\sigma(s,t|e)}{\sigma(s,t)},
\end{equation}    
where the betweenness centrality $c_B$ of a corridor e is the sum of fractions of all shortest paths between a municipality $s \in M$ and a hospital $t \in H$ that use the corridor $e$, divided by the number of all shortest paths between them (denoted by $\sigma(s,t)$). In plain words, this measures calculates how often corridors appear on the shortest paths between all pairs of municipalities and hospitals in the country at most 100km apart from one another. 

\subsection{Stress testing corridor networks}
In an effort to establish a ranking of the accessibility corridors based on their importance to hospital accessibility, we conducted two distinct stress tests of the Austrian accessibility corridor network. The first kind of test tracks the reaction of the system to the deletion of a single accessibility corridor. Specifically, we remove one link from the network and calculate the accessibility of each municipality to its closest hospital post-deletion. The ranking of the corridors was based on the resulting \acrshort{acis} for each deletion: the higher the \acrshort{acis} score a corridor receives, the higher its ranking.

\begin{figure}[ht]
	\centering
	\includegraphics[width=0.75\textwidth]{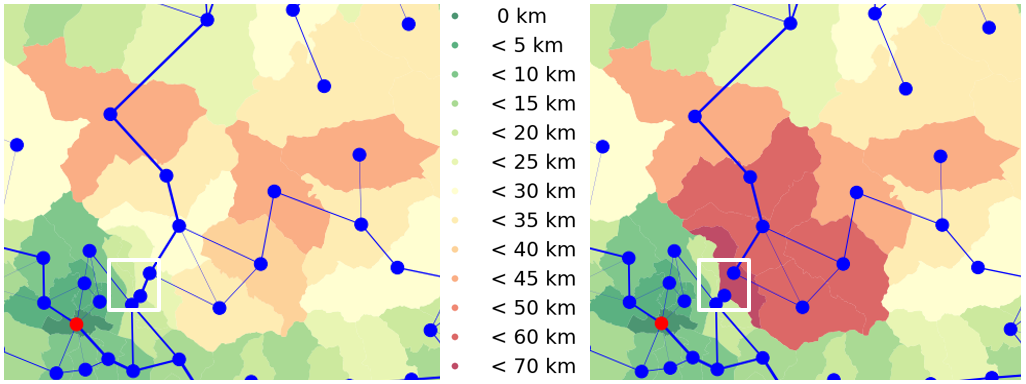}
	\caption{Here we observe how the deletion of a single corridor, highlighted in the white box, significantly impacts the hospital access of the surrounding municipalities (adapted from \cite{DRD2022}).}
	\label{fig: example}
\end{figure}

We show a concrete example of such a corridor deletion in Figure \ref{fig: example}. On the left, we color municipalities by how long they must travel to reach a hospital when the network is functioning undisturbed. Following the removal of the focal corridor, visualized on the right, people in several municipalities must travel significantly farther to reach healthcare. This would be reflected in a large \acrshort{acis} score for this corridor.

While the results of the first stress test serve as a fine approximation for the topological importance of corridors, real world events often impact roads across wider geographic areas. For instance a weather event like a snow storm could impact roads across entire regions. Even when a single corridor or road may be closed, resulting congestion may cause significant delays for travelers on nearby alternatives. 

The second stress test thus introduces neighborhood outages of roads. It measures the system's functionality after the deletion of a corridor and its neighboring corridors. This idea was initially sparked by the observation that severe weather conditions often have a widespread impact across geographic space rather than being confined to a single location. To increase the potential volatility of the stress test, we first delete the focal corridor, then with a fixed probability $p$ remove each of its immediate neighbors. This fixed probability $p$ was chosen to simulate the decreasing severity of a weather event with increasing distance from its core, which is assumed to be at the focal corridor. In particular, we ran $100$ simulations for each corridor and its neighborhood with $ p \in [0.1, 0.25, 0.5, 0.75]$. In each simulation, the hospital accessibility of the network and the distance to the closest hospital for each municipality were calculated, and the same impact measures were calculated as for the single corridor removal stress test. 
\subsection{Measuring hospital vulnerability}
While assessing the resilience of infrastructure networks and the accessibility changes caused by disturbances has been studied in previous research ~\citep{jenelius_importance_2006, xie_disrupted_2023, core:Grocery_2020,anderson_underestimated_2022}, less attention has been paid to how transportation infrastructure disturbances impact potential flows to healthcare centers. For instance, a key road closure could greatly increase the number of people going to a specific hospital as closest point of care. 

Therefore we also explored an alternative approach to measuring the impact of corridor deletions in terms of their impact on hospital catchment areas. In particular, we look for which hospitals become responsible for a significantly larger population as their closest point of care when specific corridors are closed. This allows us to measure the strain on hospitals resulting from corridor closures. To quantify this impact on hospitals, we assess how many people have to move from one hospital to another for each stress test, which can be mathematically written as:
\begin{equation}
	P_{affected} = \sum_{M \in Change} P_M,
\end{equation}
where we calculate the total affected Population $P_{affected}$ by summing over all $M \in Change$, which is the portion of municipalities that have a new closest hospital after the simulated deletion of a corridor, and $P_M$ stands for the population of municipality $M$. From this calculation, we are also able to calculate the new number of patients per hospital. Besides this, we use the different stress test results to calculate how often a hospital experiences a changing inflow due to a corridor deletion.

Through the application of these measurements, we can better understand how hospital catchment areas change due to the alteration of the accessibility corridor network. For instance, a corridor closure may change the closest hospital for a significant number of people. The changing size of the hospital catchment area thus either causes a growth or reduction in the patient flow to specific hospitals, straining or relaxing those hospitals' capacity, respectively. By examining the effect of corridor deletions on hospital catchment areas, we can derive a map of redundancy relationships between hospitals. This map allows us to report hospitals that could be more prone to sudden patient influx during crisis events which lead to corridor closures.

\section{Results}
\label{sec:results}
In this section, we present the results of our analyses. We first focus on the single corridor deletion stress test. We find that according to the ACIS measure, the impact of such deletions is highly heterogeneous: some corridors are significantly more important than the average one. We show a significant correlation between the ACIS measure and the $HA$ measure of corridor importance in this scenario. We investigate the relationship between corridor ACIS score and the number of roads in a corridor, finding highly important corridors containing very few roads. We also analyze changes in travel times. We then analyze the results of the corridor neighborhood stress test. Finally, we present two case studies in which hospitals are often or significantly impacted by corridor closures.

\subsection{Single Corridor stress test}
\begin{figure}[h]
	\centering
	\begin{subfigure}[b]{0.49\textwidth}
		\centering
		\includegraphics[width=\textwidth]{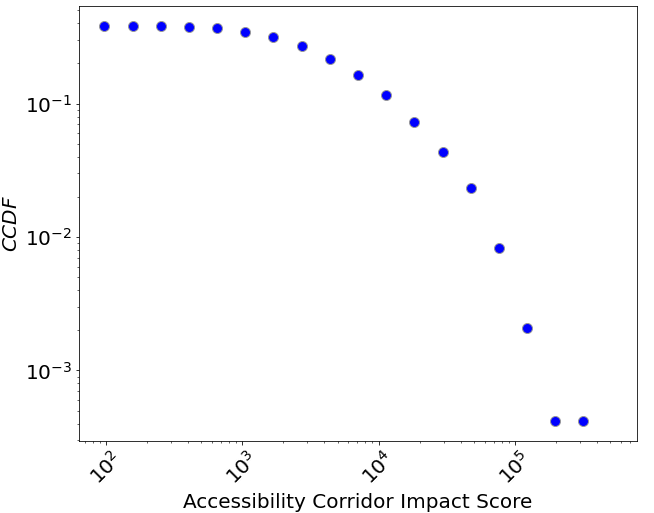}
		\caption{The CCDF of the \acrlong{acis} scores under the single accessibility corridor deletion stress test. This stress test can be used to simulate the failure of a single corridor due to weather events with geographically limited impacts, i.e., avalanche or mudflow.}
		\vspace{20pt}
		\label{fig:singleCCDF}
	\end{subfigure}
	\hfill
	\begin{subfigure}[b]{0.49\textwidth}
		\centering
		\includegraphics[width=\textwidth]{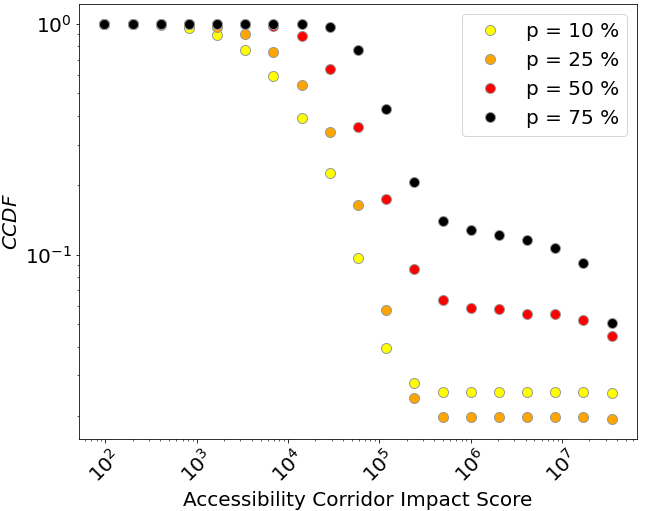}
		\caption{The CCDF of the 90th percentile from $100$ simulation of the \acrlong{acis} scores of neighborhood accessibility corridor deletion stress test, with a deletion probability $p \in [10\%, 25\%, 50\%, 75\%]$ for the neighboring streets. This stress test is used to simulate a simplified weather event, where we assume that the impact on the road network decreases with increasing distance from the core located at the focal street.}
		\label{fig:CompareCCDF}
	\end{subfigure}
	\caption{The CCDF of the \acrlong{acis} scores calculated from the deletion of Accessibility Corridors under different stress tests. In general, we observe that most corridors are not critical in providing access to hospitals but that a few are orders of magnitude more important.}
	\label{fig: CCDF}
\end{figure}

First, we found that most accessibility corridors closures have a low impact on the population, see Figure \ref{fig:singleCCDF}. In this figure we plot the complementary cumulative density function (CCDF) of the \acrlong{acis} score of each corridor. In general, the closure of any given corridor is a minor nuisance in terms of getting to a hospital. However, there are a few accessibility corridors which have a tremendous impact on the system if closed, observed in the right tail of this figure. Furthermore, the results of the neighborhood deletion stress test, see Figure \ref{fig:CompareCCDF}, suggest that locally correlated corridor closures can have an even greater impact. This is our first important result: as a few corridors are much more important than the typical one, policymakers can focus their attention on just a few parts of the (abstracted) road network. Improvements to the resilience at these key points can make a significant difference in its resilience.

\begin{figure}[h]
	\centering
	\includegraphics[width=1\textwidth]{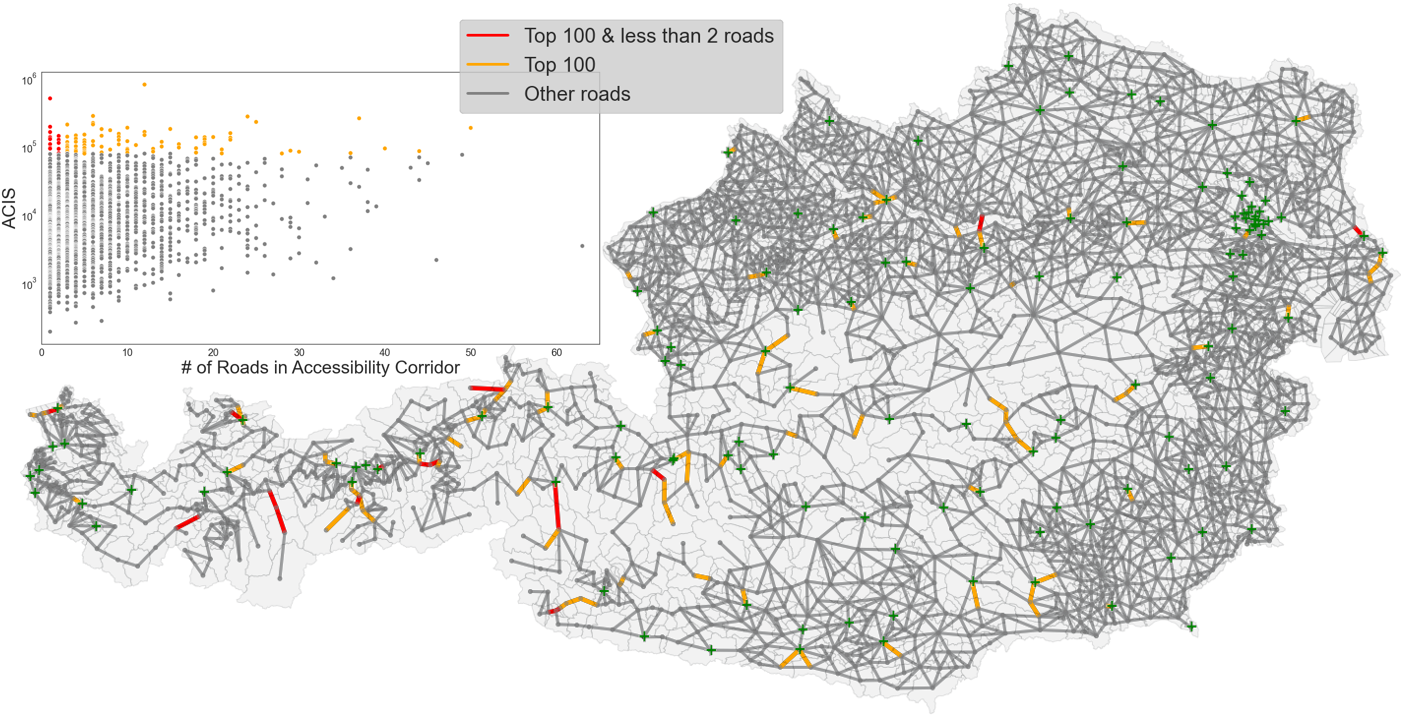}
	\caption{The top 100 most important corridors based on \acrshort{acis} under the single corridor deletion stress test. Inset: the relationship between the number of roads in a corridor and its \acrshort{acis} score. We observe critically important corridors containing few roads.}
	\label{fig: first results}
\end{figure}

A comprehensive representation of the simulation results using the \acrfull{acis} of the single corridor stress test can be found in Figure \ref{fig: first results}. To provide context for these findings, we now interpret which corridors play a crucial role according to this first stress test. In the map, we see that the highlighted corridors seem to function as connectors to otherwise isolated dead-ends to the network or as critical connectors reducing travel time between different regions. Another category of highlighted corridors are short-cuts directly connected to a hospital.

As corridors are abstractions that bundle together any number of roads between two neighboring municipalities, we look more closely at the relationship between the \acrshort{acis} ranking and the number of roads within a corridor in the inset of Figure \ref{fig: first results}. We find that there are many examples of corridors containing just a few roads and having a high \acrshort{acis}. These corridors are perhaps the most important ones to focus on: they are both systemically important and contain few local redundancies. This is especially relevant when the topography of Austria is considered as many valleys in the Alps are only connected to the rest of Austria by a single corridor. If a road like that is blocked, the access to a hospital of the municipalities in the valley is cut off.

What about the other measures of corridor importance? Under the single deletion stress test scenario, \acrshort{acis} and $HA(k)$ are highly correlated (Spearman's $\rho = 0.83$).
Edge betweenness centrality on the other hand is not significantly correlated (Spearman's $\rho = 0.09$) with the \acrshort{acis} measure, nor with the $HA(k)$ measure (Spearman's $\rho = 0.098$). As edge betweenness centrality evaluates corridor importance in terms of access to multiple hospitals, we focus on the other measures as they capture access to the closest point of care. 

\begin{figure}[ht]
	\centering
	\includegraphics[width=1\textwidth]{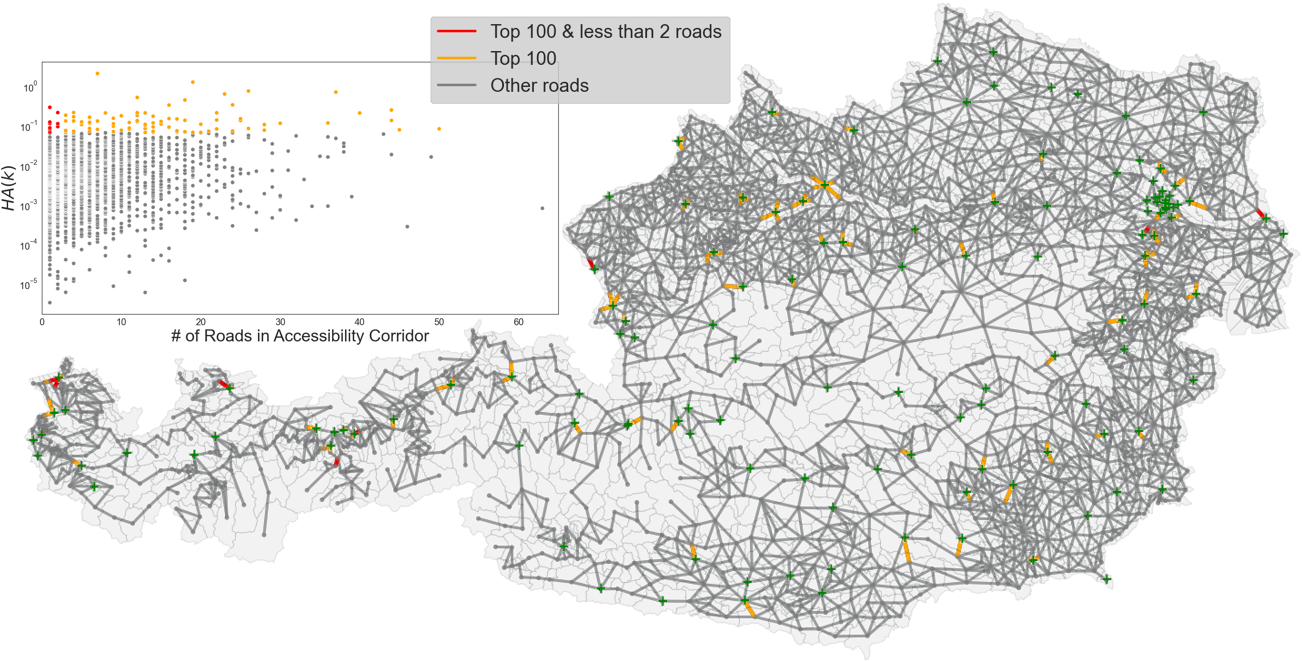}
	\caption{The top 100 most important corridors based on the accessibility corridor importance factor $HA(k)$ under the single corridor deletion stress test. Inset: the relationship between the number of roads in a corridor and its $HA(k)$ score. We observe critically important corridors containing few roads.}
	\label{fig: HA map}
\end{figure}

Upon closer examination, we found that \acrshort{acis} and $HA(k)$ do deviate significantly from each other in the most important cases. If we consider the top 100 corridors according to either ranking, this correlation turns negative (Spearman's $\rho = -0.26$). This means that the two rankings significantly diverge in terms of which corridors they consider most important. In Figure \ref{fig: HA map}, we plot the map of Austria with important corridors according to the $HA(k)$ measure highlighted. We observe that the $HA(k)$ measure tends to rank the last corridors leading directly to hospitals as the most important, while the \acrshort{acis} measure tends to emphasize corridors that appear to bridge regions. Among the top 100 $HA(k)$ ranked corridors, the average corridor contains $14$ roads, while the top 100 \acrshort{acis} ranked corridors contain on average only $11$ roads. This suggests that the \acrshort{acis} is ranking highly those corridors that bridge regions and are highly vulnerable due to their dependence on fewer road segments. In the rest of the paper, we therefore focus on the \acrshort{acis} measure.

\begin{figure}[h]
	\centering
	\begin{subfigure}[b]{0.4\textwidth}
		\centering
		\includegraphics[width=\textwidth]{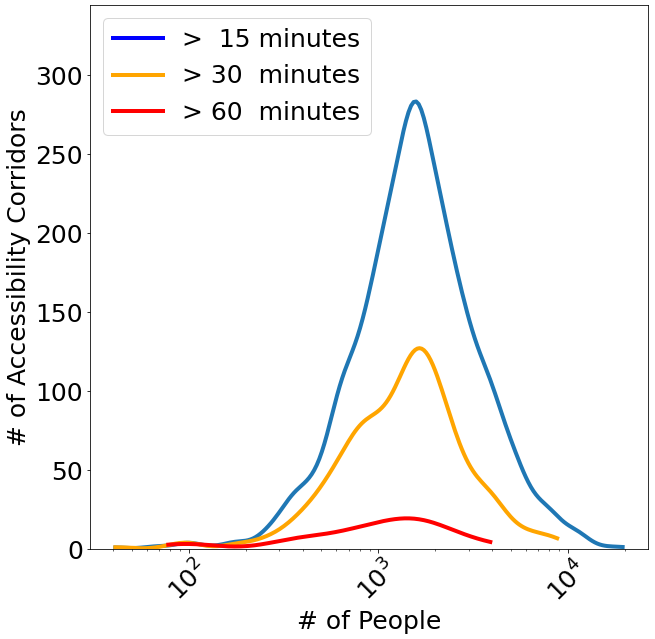}
		\captionsetup{labelformat=empty} 
		\caption{\textbf{Figure 6a}: How many accessibility corridors cause through their deletion $y$ people to drive more than $x$ minutes in total to the nearest hospital.}
		\label{fig:timechange}
	\end{subfigure}%
	\hfill
	\begin{subfigure}[b]{0.58\textwidth}
		\centering
		\renewcommand{\arraystretch}{1.37}
		\begin{tabular}{|lcc|cc|c|}
			\hline
			\multicolumn{3}{|c|}{Accessibility Corridors}                           & \multicolumn{2}{c|}{$\min(\text{Driving Time})$}                               & \multirow{2}{*}{Population} \\ \cline{1-5}
			\multicolumn{1}{|c|}{From}    & \multicolumn{1}{c|}{To}      & Roads & \multicolumn{1}{c|}{Initial} & after Del &                                                                                 \\ \hline
			\multicolumn{1}{|l|}{AT50613} & \multicolumn{1}{c|}{AT70717} & 1        & \multicolumn{1}{c|}{23,60}           & 41,40                    & 10,143                                                                        \\ \hline
			\multicolumn{1}{|l|}{AT70715} & \multicolumn{1}{c|}{AT70705} & 3        & \multicolumn{1}{c|}{16,50}           & 44,00                    & 9,914                                                                        \\ \hline
			\multicolumn{1}{|l|}{AT70705} & \multicolumn{1}{c|}{AT70704} & 3        & \multicolumn{1}{c|}{24,10}           & 44,00                    & 8,128                                                                      \\ \hline
			\multicolumn{1}{|l|}{AT70701} & \multicolumn{1}{c|}{AT70704} & 3        & \multicolumn{1}{c|}{28,10}           & 44,00                    & 6,899                                                                         \\ \hline
			\multicolumn{1}{|l|}{AT70729} & \multicolumn{1}{c|}{AT70701} & 2        & \multicolumn{1}{c|}{30,60}           & 44,00                    & 6,252                                                                     \\ \hline
			\multicolumn{1}{|l|}{AT70828} & \multicolumn{1}{c|}{AT70814} & 3        & \multicolumn{1}{c|}{2,70}            & 27,50                    & 5,696                                                                         \\ \hline
			\multicolumn{1}{|l|}{AT70203} & \multicolumn{1}{c|}{AT70825} & 1        & \multicolumn{1}{c|}{7,40}            & 29,70                    & 2,923                                                                         \\ \hline
			\multicolumn{1}{|l|}{AT70809} & \multicolumn{1}{c|}{AT70825} & 1        & \multicolumn{1}{c|}{15,70}           & 29,70                    & 2,818                                                                        \\ \hline
		\end{tabular}
		\captionsetup{skip=32pt} 
		\captionsetup{labelformat=empty} 
		\caption{\textbf{Table 6b}: Impact of the deletion of an accessibility corridor on the minimum travel time sorted by affected population, including the number of real roads in an accessibility corridor.}
		\label{fig:MuniTT}
	\end{subfigure}
	\captionsetup{labelformat=empty} 
	\label{fig: time change caused by del}
\end{figure}

To make the analysis more concrete, we report changes in driving times experienced by people following the single deletion stress test in Figure \ref{fig:timechange}. Specifically, we examine the number of additional individuals who would need to drive at least 15, 30, or 60 minutes, assuming a tempo limit of $50 \text{km}/\text{h}$, following a corridor deletion. We observe that a significant number of people would have to drive more than 15 minutes following such a deletion. As before, it is worth focusing on the extreme cases: some corridors cause thousands of people to have to drive over 60 minutes to get to a hospital. We report specific examples in Table \ref{fig:MuniTT}: the deletion of the corridor reported in the first row, which contains a single road, causes a nearly 20-minute increase in driving time for over 10,000 people. Such delays can make a significant difference in critical care outcomes. To that end we report those corridors whose deletion increases average travel time by at least five minutes in \Cref{tab: delta time difference}. Such a difference has been shown to cause statistically observable higher 30-day mortality rate in critical cases, cf.~\cite{jena_delays_2017}. For example, the deletion of the corridor in the first row of this \Cref{tab: delta time difference} leads to a mean increase of approximately $7$ minutes for more than $40,000$ people.

\begin{table}[]
	\renewcommand{\arraystretch}{1.37}
	\begin{tabular}{|cc|ccc|c|c|c|}
		\hline
		\multicolumn{2}{|c|}{Deleted Corridor}  & \multicolumn{3}{c|}{Travel Time}                                & \multirow{2}{*}{\begin{tabular}[c]{@{}c@{}}affected\\ Population\end{tabular}} & \multirow{2}{*}{\begin{tabular}[c]{@{}c@{}}affected\\ Municipalties\end{tabular}} & \multirow{2}{*}{Roads} \\ \cline{1-5}
		\multicolumn{1}{|c|}{From}    & To      & \multicolumn{1}{c|}{Old}   & \multicolumn{1}{c|}{Delta} & New   &                                                                                &                                                                                   &                        \\ \hline
		\multicolumn{1}{|c|}{AT41743} & AT41746 & \multicolumn{1}{c|}{18.48} & \multicolumn{1}{c|}{7.29}  & 25.77 & 42656 & 14 & 24                     \\ \hline
		\multicolumn{1}{|c|}{AT70926} & AT70921 & \multicolumn{1}{c|}{31.33} & \multicolumn{1}{c|}{7.89}  & 39.19 & 30791 & 21 & 6                      \\ \hline
		\multicolumn{1}{|c|}{AT70332} & AT70331 & \multicolumn{1}{c|}{25.45} & \multicolumn{1}{c|}{6.56}  & 32.07 & 27080 & 15 & 7                      \\ \hline
		\multicolumn{1}{|c|}{AT61045} & AT61053 & \multicolumn{1}{c|}{13.98} & \multicolumn{1}{c|}{8.8}   & 22.78 & 26548 & 8 & 37                     \\ \hline
		\multicolumn{1}{|c|}{AT70921} & AT70910 & \multicolumn{1}{c|}{34.83} & \multicolumn{1}{c|}{6.68}  & 41.47 & 26260 & 18 & 2                      \\ \hline
		\multicolumn{1}{|c|}{AT70331} & AT70350 & \multicolumn{1}{c|}{27.01} & \multicolumn{1}{c|}{6.5}   & 33.57 & 24850 & 14 & 4                      \\ \hline
		\multicolumn{1}{|c|}{AT70910} & AT70935 & \multicolumn{1}{c|}{36.28} & \multicolumn{1}{c|}{6.28}  & 42.52 & 24824 & 17 & 2                      \\ \hline
		\multicolumn{1}{|c|}{AT70935} & AT70923 & \multicolumn{1}{c|}{37.58} & \multicolumn{1}{c|}{9.53}  & 47.13 & 22987 & 16 & 8                      \\ \hline
		\multicolumn{1}{|c|}{AT70413} & AT70416 & \multicolumn{1}{c|}{17.0}  & \multicolumn{1}{c|}{7.63}  & 24.6  & 21291 & 6 & 4                      \\ \hline
		\multicolumn{1}{|c|}{AT31839} & AT31818 & \multicolumn{1}{c|}{19.58} & \multicolumn{1}{c|}{7.82}  & 27.38 & 18747 & 4 & 12                     \\ \hline
		\multicolumn{1}{|c|}{AT61114} & AT61108 & \multicolumn{1}{c|}{23.3}  & \multicolumn{1}{c|}{9.65}  & 32.92 & 18027 & 4 & 9                      \\ \hline
		\multicolumn{1}{|c|}{AT20923} & AT20913 & \multicolumn{1}{c|}{24.25} & \multicolumn{1}{c|}{10.45} & 34.7  & 17898 & 4 & 50                     \\ \hline
		\multicolumn{1}{|c|}{AT10706} & AT10724 & \multicolumn{1}{c|}{37.1}  & \multicolumn{1}{c|}{6.03}  & 43.1  & 17543 & 7 & 5                      \\ \hline
		\multicolumn{1}{|c|}{AT20101} & AT20402 & \multicolumn{1}{c|}{27.3}  & \multicolumn{1}{c|}{6.18}  & 33.45 & 16737 & 6 & 19                     \\ \hline
		\multicolumn{1}{|c|}{AT40101} & AT41624 & \multicolumn{1}{c|}{13.55} & \multicolumn{1}{c|}{6.7}   & 20.25 & 16731 & 4 & 9                      \\ \hline
	\end{tabular}
	\caption{Impact of the deletion of an accessibility corridor (Top 15) on the travel time with a focus on the time difference. We include the number of roads in each accessibility corridor and the affected population and report the largest impacted populations thresholded by five minutes.}
	\label{tab: delta time difference}
\end{table}

\subsection{Corridor Neighborhood stress test}
We now discuss the results of our second stress test, where we simulated the deletion of corridor neighborhoods to see the network's reaction to more significant alterations. To recapitulate, the main idea of the second stress test extend the first stress by introducing local geographic correlations in road closures, reflecting for example the broader impacts of extreme weather. Besides ranking the corridor neighborhoods, we will also compare the rankings to the initial stress test to see if the same areas are impacted.

Each instance of the second stress test also focuses on a single focal corridor. It additionally considers all neighboring corridors, removing them from the system with a probability $p$. For each focal corridor we ran 100 simulations for each $ p \in [0.1, 0.25, 0.5, 0.75]$. This approach yields a distribution of impact scores for each corridor. For each focal corridor and $p$ we considered the mean and the $90$th percentile of the \acrshort{acis} of this distribution of results.

For low values of $p$, i.e. $0.1$, the Spearman correlation between \acrshort{acis} of the single corridor deletion stress test and the neighborhood deletion stress test are high: 0.57 for the mean and 0.41 for the 90th percentile result. However, this correlation quickly drops as we consider higher likelihoods of correlated corridor failures. At $p=0.75$, the correlations drop to 0.08 for the mean and 0.03 for the 90th percentile. We may conclude this implies that when corridors are failing in a larger geographic area, as is often the case, the impact on hospital access is very different from the situation in which a single corridor is removed. 

\begin{figure}[ht]
	\centering
	\includegraphics[width=1\textwidth]{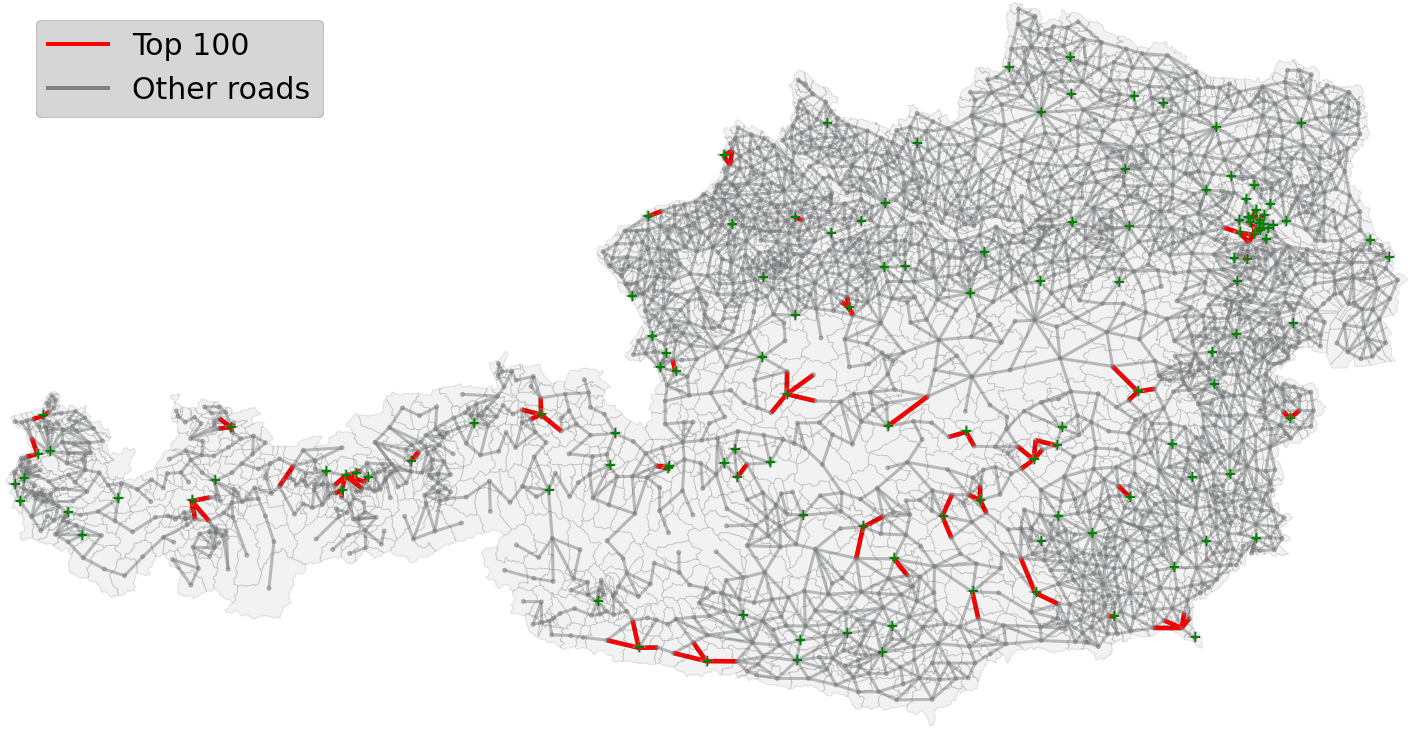}
	\caption{The Top 100 most important corridors based on ACIS under the neighborhood corridor deletion stress test with a deletion probability of $25 \%$ for the neighboring streets. We observe that there is a notable presence of  corridors directly connected to hospitals among the Top 100.}
	\label{fig: neigh results}
\end{figure}

Indeed, in \Cref{fig: neigh results}, we observe that the top 100 most important corridors according to the neighborhood deletion stress test are quite different from those under the single corridor deletion stress test when the deletion probability is increased. Calculating the overlap of the top 100 most impactful corridors of the single \acrshort{acis} with the \acrshort{acis} ranking considering a probability of $25\%$ for neighborhood deletions shows that only $15\%$ of the corridors are ranked in the top 100. This overlap is even lower for higher probabilities. Furthermore, it is apparent from this visualization that in many cases, important corridors are directly connected to a hospital forming small clusters, indirectly highlighting vulnerable neighborhoods within the corridor network as a whole.

\subsection{Hospital vulnerability}
In this section, we will study the vulnerability of hospitals based on the single corridor removal stress test. By closely examining the relations between accessibility corridor deletions and hospital catchment area shifts, we are able to derive a map of redundancies between hospitals, and to quantify which hospitals are at risk of suddenly becoming responsible for a significantly greater number of patients due to road closures. The resulting map provides detailed insight into the dynamics of patient flow between the hospitals caused by the alteration of the corridor network.

\begin{figure}
	\centering
	\includegraphics[width=1\textwidth]{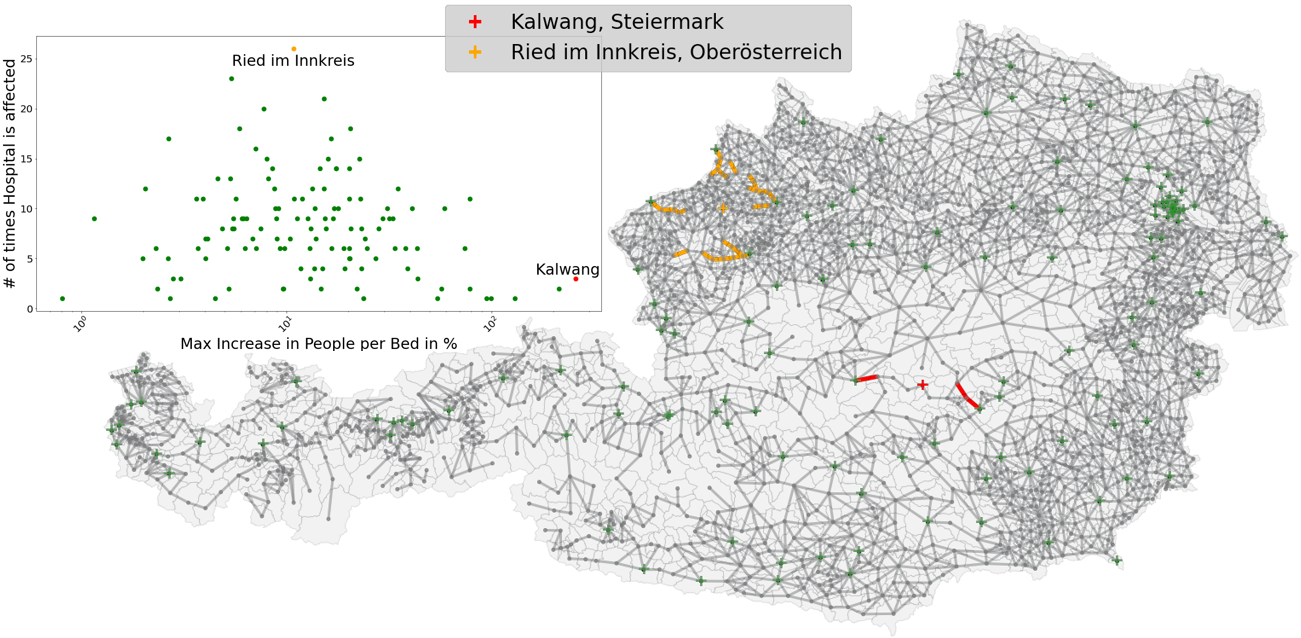}
	\caption{Exploring the impact of corridor deletions on the hospitals in the network: comparing the probability a hospital is affected to the stress tests' impact on the hospital. In this illustration, a comparison is made between a high-probability case with a moderate impact and a low probability with a high impact.}
	\label{fig: bed results}
\end{figure}

The map and scatter plot inset in Figure \ref{fig: bed results} offers a compelling portrayal of the frequency and magnitude of impacts that hospitals experience during the different stress tests. To delve deeper into the analysis, we have selected two illuminating cases that exemplify two kinds of vulnerable hospitals. Our first example, located in Kalwang, is a hospital that is only impacted by a few specific corridor closures. However, when those corridors close, the impact is extreme: with a 250\% increase in the number of people in its catchment area per bed, see Table \ref{tab:lowprobcase}. These corridors would otherwise provide access to hospitals in Rottenmann (162 beds) or Leoben (408 beds). In other words, closures of nearby corridors can lead to dramatic surges at this hospital, with a capacity of only 72 beds.

The second example, located in Ried im Innkreis, is rather potentially impacted by many corridor closures, but to a smaller degree. Over 20 different corridors can impact its catchment area, but they increase the population to bed ratio by less than 10\%. In other words, this hospital will likely often see small increases in the population for which it is the first point of service. Such small increases can nevertheless be the source of significant volatility over time in hospital admittances.

\begin{table}[b]
	\begin{tabular}{|cc|cc|cc|c|}
		\hline
		\multicolumn{2}{|c|}{People per Bed}           & \multicolumn{2}{c|}{Change per Bed} & \multicolumn{2}{c|}{Accessibility Corridor}        & \multicolumn{1}{l|}{\multirow{2}{*}{\begin{tabular}[c]{@{}l@{}}Affected \\ Population\end{tabular}}} \\ \cline{1-6}
		\multicolumn{1}{|c|}{Initial} & After Deletion & \multicolumn{1}{c|}{in \%} & People & \multicolumn{1}{c|}{Deleted Corridor}   & \# Roads & \multicolumn{1}{l|}{}                                                                                \\ \hline
		\multicolumn{1}{|c|}{183.56}  & 203.40         & \multicolumn{1}{c|}{10.81} & 19.83  & \multicolumn{1}{c|}{(AT41743, AT41746)} & 24       & 9 787                                                                                                 \\ \hline
		\multicolumn{1}{|c|}{183.56}  & 196.04         & \multicolumn{1}{c|}{6.80}  & 12.48  & \multicolumn{1}{c|}{(AT41426, AT41402)} & 3        & 5 203                                                                                                 \\ \hline
		\multicolumn{1}{|c|}{183.56}  & 195.31         & \multicolumn{1}{c|}{6.40}  & 11.74  & \multicolumn{1}{c|}{(AT40419, AT40418)} & 5        & 4 897                                                                                                 \\ \hline
		\multicolumn{1}{|c|}{183.56}  & 195.28         & \multicolumn{1}{c|}{6.38}  & 11.71  & \multicolumn{1}{c|}{(AT41716, AT41743)} & 2        & 4 884                                                                                                 \\ \hline
		\multicolumn{1}{|c|}{183.56}  & 195.28         & \multicolumn{1}{c|}{6.38}  & 11.71  & \multicolumn{1}{c|}{(AT41716, AT41709)} & 15       & 4 884                                                                                                 \\ \hline
		\multicolumn{1}{|c|}{183.56}  & 193.96         & \multicolumn{1}{c|}{5.66}  & 10.39  & \multicolumn{1}{c|}{(AT40830, AT40808)} & 4        & 4 334                                                                                                 \\ \hline
		\multicolumn{1}{|c|}{183.56}  & 191.69         & \multicolumn{1}{c|}{4.42}  & 8.12   & \multicolumn{1}{c|}{(AT41711, AT41743)} & 8        & 3 387                                                                                                 \\ \hline
		\multicolumn{1}{|c|}{183.56}  & 191.69         & \multicolumn{1}{c|}{4.42}  & 8.12   & \multicolumn{1}{c|}{(AT41747, AT41711)} & 9        & 3 387                                                                                                 \\ \hline
		\multicolumn{1}{|c|}{183.56}  & 190.07         & \multicolumn{1}{c|}{3.54}  & 6.50   & \multicolumn{1}{c|}{(AT41422, AT41418)} & 15       & 2 712                                                                                                 \\ \hline
	\end{tabular}
	\caption{The ten cases in this table exemplify the significant impact on the hospital in Ried im Innkreis, Oberösterreich, with a focus on the change of people per bed.}
	\label{tab: high prob case}
\end{table}

\begin{table}[h]
	\begin{tabular}{|cc|cc|cc|c|}
		\hline
		\multicolumn{2}{|c|}{People per Bed}           & \multicolumn{2}{c|}{Change per Bed} & \multicolumn{2}{c|}{Accessibility Corridor}        & \multicolumn{1}{l|}{\multirow{2}{*}{\begin{tabular}[c]{@{}l@{}}Affected \\ Population\end{tabular}}} \\ \cline{1-6}
		\multicolumn{1}{|c|}{Initial} & After Deletion & \multicolumn{1}{c|}{in \%} & People & \multicolumn{1}{c|}{Deleted Corridor}   & \# Roads & \multicolumn{1}{l|}{}                                                                                \\ \hline
		\multicolumn{1}{|c|}{83.32}   & 298.23         & \multicolumn{1}{c|}{257.96} & 214.92 & \multicolumn{1}{c|}{(AT61114, AT61108)} & 9   &   15 689  \\ \hline
		\multicolumn{1}{|c|}{83.32}   & 298.23         & \multicolumn{1}{c|}{257.96} & 214.92 & \multicolumn{1}{c|}{(AT61114, AT61120)} & 16   &   15 689 \\ \hline
		\multicolumn{1}{|c|}{83.32}   & 134.27         & \multicolumn{1}{c|}{61.16}  & 50.96  & \multicolumn{1}{c|}{(AT61263, AT61247)} & 10    &  3 720 \\ \hline
	\end{tabular}
	\caption{The following cases exemplify the significant impact on the hospital in Kalwang, Steiermark, with a focus on the change of people per bed.}
	\label{tab:lowprobcase}
\end{table}

Even though the latter is an extreme case where the closure of accessibility corridors blocks the way to big hospitals and puts a lot of strain on a small hospital, it shows how small changes can have big effects. Both cases show how our developed method can be used to refine the analysis and offer another perspective to the stress test. 

\section{Conclusion and Future Research}
\label{sec:concl}

In this paper, we show the resilience of a population's access to healthcare can be meaningfully analyzed and quantified based on stress tests of road-based transportation networks. These stress tests can provide meaningful insights into the dependencies between different systems, in this case of the transportation system and hospitals. By ranking corridors based on their importance in terms of hospital access, we can identify corridors of interest for policymakers seeking to allocate limited resources. In particular, we show that there are high impact corridors containing few roads which provide access to emergency care for many people. We also show that some hospitals are vulnerable to sudden surges of people they are responsible for.

Based on these results further investigations can be conducted, for instance by including the area around the corridors of interest. This would lead to guidelines to improve the underlying network's resilience and general access to health care. Analyzing the corridors' neighborhoods is another step from the abstract model to real-world scenarios where natural disasters impact whole regions. 

Our results show that certain road segments and corridors play a pivotal role in access to hospitals in Austria. The disruption of these roads during crisis scenarios can have a significant impact on travel time to hospitals for large numbers of people. Specific municipalities are especially vulnerable to the closure of specific road corridors. Hospitals can also be affected: a road closure can change which hospital is closest for large numbers of people. The skewed importance of some corridors highlights vulnerabilities but also gives policymakers something to focus on.

Compared to previous work, we focus on the specific problem of accessibility of hospitals in our stress test, using a generated measurement as well as adapting an appropriate accessibility measure \citep{core:martin_assessing_2021}. As we are interested in hospital accessibility, which is a local network problem, this derived measure of road importance is more appropriate than a global measure such as edge betweenness centrality. Further comparison shows that the measurement generated in this paper is a better fit for our problem since the accessibility measure from \citep{core:martin_assessing_2021} over-emphasizes city size for our application. In the scenario described in the paper, the size of a hospital doesn't add to its attractiveness, and the focus is solely on the accessibility of medical care. If the size of the target is important to the simulation, for example, the size of a city or the number of hospital beds, an adaptation of \citep{core:martin_assessing_2021} is more practical. However, in this simulation, we assume that fast access to health care is crucial and not influenced by the size of the hospital. 

Our work also has policy implications for the healthcare sector. Past research has demonstrated that hospital capacity has a tipping-point in terms of care quality: when occupancy exceeds a critical level of capacity, mortality outcomes worsen significantly \cite{kuntz2015stress}. Our work shows that such surges can occur due to the outcomes in another complex system. Flexible staffing and pooled capacity across hospitals, effective policies recommended in this previous work, should take into account how exogeneous shocks influencing transport networks can create surges and limit the effectiveness of these interventions.

Rather than vulnerability to specific events (i.e., floods \citep{core:van_ginkel_will_2022}), we consider abstract road closures. By coarse-graining the Austrian road network, we can run a more thorough stress test on a country-wide network. Both of these simplifications enable us to easily identify network sections of interest. More fine-grained versions of these sections can be further investigated in future work using more realistic stress tests. This would be especially tractable if policymakers wished to zoom in on a specific region or part of the country.

Our study has several limitations. Roads in different parts of the country may be more or less vulnerable to closure, given factors like local weather and altitude. Future work should consider historical weather patterns and their correlation with road closures. More realistic stress tests can be developed in this way. Some but not all of the road corridors we highlight pass through extremely rugged terrain (i.e. the Alps). Creating redundancies in this context may be very expensive. In such areas it may be more useful to create redundancies at highly impact hospitals.

Furthermore, our results show that overlaying the \acrshort{acis} measurement with the number of real-world roads in a corridor can help to understand the implications of a corridor closure better. Therefore, we propose to update the introduce \acrshort{acis} method by including a factor that takes the number of real-world roads into account. 

More generally, our approach to stress-testing road networks can be applied in a variety of contexts. For example, rather than measuring the accessibility of hospitals from population centers, we may measure accessibility of population centers from firefighting stations. Indeed, many critical services rely on the functionality of transportation systems like the road network. Social, demographic, and environmental factors suggest that these systems will only experience greater strain in the coming decades. Whether due to climate change, large-scale migration, or the aging of the population, resilience and robustness of these services as a function of the systems they depend on will merit increasing scrutiny. Integrating various forms of data, for example, in a knowledge graph designed for crisis management \cite{anjomshoaa2023towards}, can greatly expand the potential scope of our simulations. The knowledge graph can be used in future works to determine additional risk factors for roads due to overlapping networks, e.g., river networks which can increase the risk to a road located close by or crossing the river. By adding these risk factors to the simplified network, the simulation can be adapted by updating the probabilities, and consequently, the relevance of the findings to real-world situations can be improved.
\appendix

\printcredits

\bibliographystyle{plainnat}
\bibliography{cas-refs}

\begin{thebibliography}{31}
\providecommand{\natexlab}[1]{#1}
\providecommand{\url}[1]{\texttt{#1}}
\expandafter\ifx\csname urlstyle\endcsname\relax
  \providecommand{\doi}[1]{doi: #1}\else
  \providecommand{\doi}{doi: \begingroup \urlstyle{rm}\Url}\fi

\bibitem[Anderson et~al.(2022)Anderson, Kiddle, and
  Logan]{anderson_underestimated_2022}
M.~J. Anderson, D.~A.~F. Kiddle, and T.~M. Logan.
\newblock The underestimated role of the transportation network: Improving
  disaster \& community resilience.
\newblock \emph{Transportation Research Part D: Transport and Environment},
  106:\penalty0 103218, 2022.
\newblock ISSN 1361-9209.
\newblock \doi{10.1016/j.trd.2022.103218}.
\newblock URL
  \url{https://www.sciencedirect.com/science/article/pii/S1361920922000487}.

\bibitem[Anjomshoaa et~al.(2023)Anjomshoaa, Schuster, Wachs, and
  Polleres]{anjomshoaa2023towards}
Amin Anjomshoaa, Hannah Schuster, Johannes Wachs, and Axel Polleres.
\newblock Towards crisis response and intervention using knowledge graphs-crisp
  case study.
\newblock In \emph{International Conference on Advanced Information Systems
  Engineering}, pages 67--73. Springer, 2023.

\bibitem[Antoniou and Tsompa(2008)]{antoniou_statistical_2008}
I.~E. Antoniou and E.~T. Tsompa.
\newblock Statistical analysis of weighted networks.
\newblock \emph{Discrete Dynamics in Nature and Society}, 2008:\penalty0 1--16,
  2008.
\newblock ISSN 1026-0226, 1607-887X.
\newblock \doi{10.1155/2008/375452}.
\newblock URL \url{http://www.hindawi.com/journals/ddns/2008/375452/}.

\bibitem[Battiston et~al.(2016)Battiston, Farmer, Flache, Garlaschelli,
  Haldane, Heesterbeek, Hommes, Jaeger, May, and
  Scheffer]{battiston2016complexity}
Stefano Battiston, J~Doyne Farmer, Andreas Flache, Diego Garlaschelli, Andrew~G
  Haldane, Hans Heesterbeek, Cars Hommes, Carlo Jaeger, Robert May, and Marten
  Scheffer.
\newblock Complexity theory and financial regulation.
\newblock \emph{Science}, 351\penalty0 (6275):\penalty0 818--819, 2016.

\bibitem[Bok{\'a}nyi and Hann{\'a}k(2020)]{bokanyi2020understanding}
Eszter Bok{\'a}nyi and Anik{\'o} Hann{\'a}k.
\newblock Understanding inequalities in ride-hailing services through
  simulations.
\newblock \emph{Scientific reports}, 10\penalty0 (1):\penalty0 6500, 2020.

\bibitem[Bíl et~al.(2015)Bíl, Vodák, Kubeček, Bílová, and
  Sedoník]{BIL201590}
Michal Bíl, Rostislav Vodák, Jan Kubeček, Martina Bílová, and Jiří
  Sedoník.
\newblock Evaluating road network damage caused by natural disasters in the
  czech republic between 1997 and 2010.
\newblock \emph{Transportation Research Part A: Policy and Practice},
  80:\penalty0 90--103, 2015.
\newblock ISSN 0965-8564.
\newblock \doi{https://doi.org/10.1016/j.tra.2015.07.006}.
\newblock URL
  \url{https://www.sciencedirect.com/science/article/pii/S0965856415001883}.

\bibitem[Diem et~al.(2020)Diem, Pichler, and Thurner]{diem2020minimal}
Christian Diem, Anton Pichler, and Stefan Thurner.
\newblock What is the minimal systemic risk in financial exposure networks?
\newblock \emph{Journal of Economic Dynamics and Control}, 116:\penalty0
  103900, 2020.

\bibitem[Doyle(2002)]{carlson2002complexity}
John Doyle.
\newblock Complexity and robustness.
\newblock \emph{Proceedings of the national academy of sciences}, 99\penalty0
  (suppl\_1):\penalty0 2538--2545, 2002.

\bibitem[Goldbeck et~al.(2019)Goldbeck, Angeloudis, and
  Ochieng]{goldbeck_resilience_2019}
Nils Goldbeck, Panagiotis Angeloudis, and Washington~Y. Ochieng.
\newblock Resilience assessment for interdependent urban infrastructure systems
  using dynamic network flow models.
\newblock \emph{Reliability Engineering \& System Safety}, 188:\penalty0
  62--79, 2019.
\newblock ISSN 0951-8320.
\newblock \doi{10.1016/j.ress.2019.03.007}.
\newblock URL
  \url{https://www.sciencedirect.com/science/article/pii/S0951832018308937}.

\bibitem[Hackl(2019)]{hackl_risk_2019}
Jürgen Hackl.
\newblock \emph{Risk Assessments of Complex Infrastructure Systems Considering
  Spatial and Temporal Aspects}.
\newblock phdthesis, {ETH} Zurich, 2019.
\newblock URL \url{http://hdl.handle.net/20.500.11850/345378}.
\newblock Artwork Size: 251 p. Medium: application/pdf Pages: 251 p.

\bibitem[Hackl et~al.(2018)Hackl, Lam, Heitzler, Adey, and
  Hurni]{hackl_estimating_2018}
Jürgen Hackl, Juan~Carlos Lam, Magnus Heitzler, Bryan~T. Adey, and Lorenz
  Hurni.
\newblock Estimating network related risks: A methodology and an application in
  the transport sector.
\newblock \emph{Natural Hazards and Earth System Sciences}, 18\penalty0
  (8):\penalty0 2273--2293, 2018.
\newblock ISSN 1684-9981.
\newblock \doi{10.5194/nhess-18-2273-2018}.
\newblock URL \url{https://nhess.copernicus.org/articles/18/2273/2018/}.

\bibitem[Jena et~al.(2017)Jena, Mann, Wedlund, and Olenski]{jena_delays_2017}
Anupam~B. Jena, N.~Clay Mann, Leia~N. Wedlund, and Andrew Olenski.
\newblock {Delays in Emergency Care and Mortality during Major U.S. Marathons}.
\newblock \emph{New England Journal of Medicine}, 376\penalty0 (15):\penalty0
  1441--1450, 04 2017.
\newblock ISSN 0028-4793, 1533-4406.
\newblock \doi{10.1056/NEJMsa1614073}.
\newblock URL \url{http://www.nejm.org/doi/10.1056/NEJMsa1614073}.

\bibitem[Jenelius et~al.(2006)Jenelius, Petersen, and
  Mattsson]{jenelius_importance_2006}
Erik Jenelius, Tom Petersen, and Lars-Göran Mattsson.
\newblock Importance and exposure in road network vulnerability analysis.
\newblock \emph{Transportation Research Part A: Policy and Practice},
  40\penalty0 (7):\penalty0 537--560, 2006.
\newblock ISSN 09658564.
\newblock \doi{10.1016/j.tra.2005.11.003}.
\newblock URL
  \url{https://linkinghub.elsevier.com/retrieve/pii/S096585640500162X}.

\bibitem[Kaleta et~al.(2022)Kaleta, Lasser, Dervic, Yang, Sorger, Lo~Sardo,
  Thurner, Kautzky-Willer, and Klimek]{kaleta2022stress}
Michaela Kaleta, Jana Lasser, Elma Dervic, Liuhuaying Yang, Johannes Sorger,
  D~Ruggiero Lo~Sardo, Stefan Thurner, Alexandra Kautzky-Willer, and Peter
  Klimek.
\newblock {Stress-testing the resilience of the Austrian healthcare system
  using agent-based simulation}.
\newblock \emph{Nature Communications}, 13\penalty0 (1):\penalty0 4259, 2022.

\bibitem[Kuntz et~al.(2015)Kuntz, Mennicken, and Scholtes]{kuntz2015stress}
Ludwig Kuntz, Roman Mennicken, and Stefan Scholtes.
\newblock Stress on the ward: Evidence of safety tipping points in hospitals.
\newblock \emph{Management Science}, 61\penalty0 (4):\penalty0 754--771, 2015.

\bibitem[Lin et~al.(2023)Lin, Xu, and Xie]{lin_location_2023}
Hongzhi Lin, Min Xu, and Chi Xie.
\newblock Location and capacity planning for preventive healthcare facilities
  with congestion effects.
\newblock \emph{Journal of Industrial and Management Optimization}, 19\penalty0
  (4):\penalty0 3044--3059, 2023.
\newblock ISSN 1547-5816.
\newblock \doi{10.3934/jimo.2022076}.
\newblock URL
  \url{https://www.aimsciences.org/en/article/doi/10.3934/jimo.2022076}.
\newblock Publisher: Journal of Industrial and Management Optimization.

\bibitem[Liu et~al.(2022)Liu, Li, Ma, Szymanski, Stanley, and
  Gao]{core:liu_network_2022}
Xueming Liu, Daqing Li, Manqing Ma, Boleslaw~K. Szymanski, H~Eugene Stanley,
  and Jianxi Gao.
\newblock Network resilience.
\newblock \emph{Physics Reports}, 971:\penalty0 1--108, 2022.
\newblock ISSN 03701573.
\newblock \doi{10.1016/j.physrep.2022.04.002}.
\newblock URL
  \url{https://linkinghub.elsevier.com/retrieve/pii/S0370157322001211}.

\bibitem[Lo~Sardo et~al.(2019)Lo~Sardo, Thurner, Sorger, Duftschmid, Endel, and
  Klimek]{lo2019quantification}
Donald~Ruggiero Lo~Sardo, Stefan Thurner, Johannes Sorger, Georg Duftschmid,
  Gottfried Endel, and Peter Klimek.
\newblock Quantification of the resilience of primary care networks by stress
  testing the health care system.
\newblock \emph{Proceedings of the National Academy of Sciences}, 116\penalty0
  (48):\penalty0 23930--23935, 2019.

\bibitem[Martín et~al.(2021)Martín, Ortega, Cuevas-Wizner, Ledda, and
  De~Montis]{core:martin_assessing_2021}
Belén Martín, Emilio Ortega, Rodrigo Cuevas-Wizner, Antonio Ledda, and Andrea
  De~Montis.
\newblock Assessing road network resilience: An accessibility comparative
  analysis.
\newblock \emph{Transportation Research Part D: Transport and Environment},
  95:\penalty0 102851, 2021.
\newblock ISSN 1361-9209.
\newblock \doi{10.1016/j.trd.2021.102851}.
\newblock URL
  \url{https://www.sciencedirect.com/science/article/pii/S1361920921001541}.

\bibitem[Mattsson and Jenelius(2015)]{MATTSSON201516}
Lars-Göran Mattsson and Erik Jenelius.
\newblock Vulnerability and resilience of transport systems – a discussion of
  recent research.
\newblock \emph{Transportation Research Part A: Policy and Practice},
  81:\penalty0 16--34, 2015.
\newblock ISSN 0965-8564.
\newblock \doi{https://doi.org/10.1016/j.tra.2015.06.002}.
\newblock URL
  \url{https://www.sciencedirect.com/science/article/pii/S0965856415001603}.
\newblock Resilience of Networks.

\bibitem[Mukherji et~al.(2023)Mukherji, Thorne, Cheung, Connors, Garschagen,
  Geden, Hayward, Simpson, Totin, Blok, Eriksen, Fischer, Garner, Guivarch,
  Haasnoot, Hermans, Ley, Lewis, Nicholls, Niamir, Szopa, Trewin, Howden,
  Méndez, Pereira, Pichs, Rose, Saheb, Sánchez, Xiao, and
  Yassaa]{core:mukherji_synthesis_nodate}
Aditi Mukherji, Peter Thorne, William W~L Cheung, Sarah~L Connors, Matthias
  Garschagen, Oliver Geden, Bronwyn Hayward, Nicholas~P Simpson, Edmond Totin,
  Kornelis Blok, Siri Eriksen, Erich Fischer, Gregory Garner, Céline Guivarch,
  Marjolijn Haasnoot, Tim Hermans, Debora Ley, Jared Lewis, Zebedee Nicholls,
  Leila Niamir, Sophie Szopa, Blair Trewin, Mark Howden, Carlos Méndez, Joy
  Pereira, Ramón Pichs, Steven~K Rose, Yamina Saheb, Roberto Sánchez, Cunde
  Xiao, and Noureddine Yassaa.
\newblock {SYNTHESIS} {REPORT} {OF} {THE} {IPCC} {SIXTH} {ASSESSMENT} {REPORT}
  ({AR}6), 2023.
\newblock URL \url{https://www.ipcc.ch/report/sixth-assessment-report-cycle/}.

\bibitem[Murata and Matsuda(2013)]{core:murata_association_2013}
Atsuhiko Murata and Shinya Matsuda.
\newblock {Association Between Ambulance Distance to Hospitals and Mortality
  from Acute Diseases in Japan: National Database Analysis}.
\newblock \emph{Journal of Public Health Management and Practice}, 19\penalty0
  (5):\penalty0 E23--E28, 2013.
\newblock ISSN 1078-4659.
\newblock \doi{10.1097/PHH.0b013e31828b7150}.
\newblock URL \url{https://journals.lww.com/00124784-201309000-00023}.

\bibitem[Rodrigue et~al.(2020)Rodrigue, Slack, Comtois, Anderson, Bowen,
  Ducruet, Notteboom, and Shaw]{rodrigue_transportation_2020}
Jean-Paul Rodrigue, Brian Slack, Claude Comtois, William Anderson, John Bowen,
  Cesar Ducruet, Theo Notteboom, and Shih-Lung Shaw.
\newblock Transportation and geography.
\newblock In Jean-Paul Rodrigue, editor, \emph{The Geography of Transport
  Systems}, chapter~1. Hofstra University, Department of Global Studies \&
  Geography, https://transportgeography.org, 2020.

\bibitem[Schneider et~al.(2011)Schneider, Moreira, Andrade, Havlin, and
  Herrmann]{schneider_mitigation_2011}
Christian~M. Schneider, André~A. Moreira, José~S. Andrade, Shlomo Havlin, and
  Hans~J. Herrmann.
\newblock Mitigation of malicious attacks on networks.
\newblock \emph{Proceedings of the National Academy of Sciences}, 108\penalty0
  (10):\penalty0 3838--3841, 2011.
\newblock \doi{10.1073/pnas.1009440108}.
\newblock URL \url{https://www.pnas.org/doi/full/10.1073/pnas.1009440108}.
\newblock Publisher: Proceedings of the National Academy of Sciences.

\bibitem[Schueller and Wachs(2022)]{schueller2022modeling}
Willam Schueller and Johannes Wachs.
\newblock Modeling interconnected social and technical risks in open source
  software ecosystems.
\newblock \emph{arXiv preprint arXiv:2205.04268}, 2022.

\bibitem[Schueller et~al.(2022)Schueller, Diem, Hinterplattner, Stangl,
  Conrady, Gerschberger, and Thurner]{schueller_propagation_2022}
William Schueller, Christian Diem, Melanie Hinterplattner, Johannes Stangl,
  Beate Conrady, Markus Gerschberger, and Stefan Thurner.
\newblock Propagation of disruptions in supply networks of essential goods: A
  population-centered perspective of systemic risk, 2022.
\newblock URL \url{https://papers.ssrn.com/abstract=4022513}.

\bibitem[Schuster et~al.(2022)Schuster, Wachs, and Polleres]{DRD2022}
Hannah Schuster, Johannes Wachs, and Axel Polleres.
\newblock Municipality access to hospitals via road networks.
\newblock In \emph{Konferenzband der Disaster Research Days 2022}, pages
  53--58. Disaster Competence Network Austria, 2022.
\newblock {ISBN}: 978-3-900397-04-3.

\bibitem[T{\'o}th et~al.(2022)T{\'o}th, Elekes, Whittle, Lee, and
  Kogler]{toth2022technology}
Gerg{\H{o}} T{\'o}th, Zolt{\'a}n Elekes, Adam Whittle, Changjun Lee, and
  Dieter~F Kogler.
\newblock Technology network structure conditions the economic resilience of
  regions.
\newblock \emph{Economic Geography}, 98\penalty0 (4):\penalty0 355--378, 2022.

\bibitem[van Ginkel et~al.(2022)van Ginkel, Koks, de~Groen, Nguyen, and
  Alfieri]{core:van_ginkel_will_2022}
Kees~C.H. van Ginkel, Elco~E. Koks, Frederique de~Groen, Viet~Dung Nguyen, and
  Lorenzo Alfieri.
\newblock Will river floods ‘tip’ european road networks? a robustness
  assessment.
\newblock \emph{Transportation Research Part D: Transport and Environment},
  108:\penalty0 103332, 2022.
\newblock ISSN 13619209.
\newblock \doi{10.1016/j.trd.2022.103332}.
\newblock URL
  \url{https://linkinghub.elsevier.com/retrieve/pii/S1361920922001602}.

\bibitem[Wiśniewski et~al.(2020)Wiśniewski, Borowska-Stefańska, Kowalski,
  and Sapińska]{core:Grocery_2020}
Szymon Wiśniewski, Marta Borowska-Stefańska, Michał Kowalski, and Paulina
  Sapińska.
\newblock Vulnerability of the accessibility to grocery shopping in the event
  of flooding.
\newblock \emph{Transportation Research Part D: Transport and Environment},
  87:\penalty0 102510, 2020.
\newblock ISSN 1361-9209.
\newblock \doi{https://doi.org/10.1016/j.trd.2020.102510}.
\newblock URL
  \url{https://www.sciencedirect.com/science/article/pii/S1361920920306970}.

\bibitem[Xie et~al.(2023)Xie, Bao, and Chen]{xie_disrupted_2023}
Chi Xie, Zhaoyao Bao, and Anthony Chen.
\newblock Disrupted transportation networks under different information
  availability and stochasticity situations.
\newblock \emph{Transportation Research Part C: Emerging Technologies},
  150:\penalty0 104097, 2023.
\newblock ISSN 0968-090X.
\newblock \doi{10.1016/j.trc.2023.104097}.
\newblock URL
  \url{https://www.sciencedirect.com/science/article/pii/S0968090X23000864}.

\end{thebibliography}


\end{document}